# On the relationship between interdisciplinarity and scientific impact

Vincent Larivière and Yves Gingras
Observatoire des sciences et des technologies (OST)
Centre interuniversitaire de recherche sur la science et la technologie (CIRST)
Université du Québec à Montréal
CP 8888, Succursale Centre-ville
 Montréal, Québec, H3C 3P8
[lariviere.vincent; gingras.yves]@uqam.ca

**Abstract**
This paper analyzes the effect of interdisciplinarity on the scientific impact of individual papers. Using all the papers published in Web of Science in 2000, we define the degree of interdisciplinarity of a given paper as the percentage of its cited references made to journals of other disciplines. We show that, although for all disciplines combined there is no clear correlation between the level of interdisciplinarity of papers and their citation rates, there are nonetheless some disciplines in which a higher level of interdisciplinarity is related to a higher citation rates. For other disciplines, citations decline as interdisciplinarity grows. One characteristic is visible in all disciplines: highly disciplinary and highly interdisciplinary papers have a low scientific impact. This suggests that there might be an optimum of interdisciplinarity beyond which the research is too dispersed to find its niche and under which it is too mainstream to have high impact. Finally, the relationship between interdisciplinarity and scientific impact is highly determined by the citation characteristics of the disciplines involved: papers citing citation intensive disciplines are more likely to be cited by those disciplines and, hence, obtain higher citation scores than papers citing non citation intensive disciplines.

**Introduction**
Over the last 40 years and especially since the publication of Gibbons *et al.* (1994) seminal book on the new production of knowledge, interdisciplinarity (and its analogous concepts of transdisciplinarity, multidisciplinarity, crossdisciplinarity, etc.) has been regarded as something positive which should be encouraged. In this respect, it has generated a great deal of theoretical and empirical literature (see, among others, the collective book edited by Weingart and Stehr, 2000), as well as several national (COSEPUP, 2004) and international (OECD, 1998) policy reports. One of the issues surrounding interdisciplinarity often discussed is its effect on the scientific impact of papers. Though this question has already been analyzed, especially in the context of the British RAE—where the question whether researchers involved in interdisciplinary research should be evaluated differently from those doing disciplinary research was raised—, these studies are usually based on a small sample of disciplines, university departments or journals. For instance, measuring interdisciplinarity at the level of journals, Levitt and Thelwall (2008) found that, in the natural and health sciences, multidisciplinary research (defined as papers published in journals to which more than one discipline has been attributed) have less scientific impact than disciplinary research. In the social sciences, both types of research obtain similar citation rates. Using the interdisciplinarity of cited references amongst Thomson *Current Contents* 88 subject categories and discipline-normalized citation counts of two research-intensive UK universities (N=37,000 papers), Adams, Jackson and Marshall (2007) found that the most interdisciplinary articles were in fact as much cited as the average article. They also found that the cited references of the most cited papers had average levels of interdisciplinarity. Finally, using physics research programs in the Netherlands as a case study, Rinia, van Leeuwen and van Raan (2002) have shown that absolute bibliometric indicators are indeed biased against interdisciplinary research. In particular, programs at the periphery of physics receive lower scores on non-normalized bibliometric indicators such as absolute number of citations and impact factors of journals in which they publish. Relative bibliometric indicators, however, correct for this bias. Taken together, these divergent results are likely a reflection of the different operational definitions of interdisciplinarity used in these studies.

The goal of this paper is to revisit the relationship between interdisciplinarity and scientific impact by compiling bibliometric data for all disciplines at the paper level on the percentage of references made to disciplines other than the one of the citing paper. Following the OECD's (1998) definitions, this paper defines interdisciplinarity as the integration of different disciplines working on a common object. In other words, in opposition to multidisciplinarity, where researchers from different disciplines work on the same topic without much integration, this paper looks at how articles from a given discipline cite articles from other disciplines and, thus, integrate their knowledge. Each article thus obtains a "degree" of interdisciplinarity based on its percentage of references made to papers of other disciplines. We then measure the relation between the degree of interdisciplinarity thus defined and the scientific impact of the citing papers.

The next section of this paper details the method used. It is followed by the presentation of the results and by a discussion and conclusion.

**Methods**
When measured using bibliometric methods, the concept of interdisciplinarity has been operationalized in different manners. However, most studies follow the method used by Porter and Chubin (1985), which measure the degree of interdisciplinarity of a paper by using the percentage of citations received by the paper from a different discipline or specialty or the percentage of the references it contain from a different discipline or specialty. Note that in the first case interdisciplinarity is defined through the practice of the authors of the article who decide what to refer to whereas in the second case interdisciplinarity is defined through the uses of the paper by the other disciplines who cite it. This indicator is indeed very similar to that used at a more micro-level by Tomov and Mutafov (1996) for andrology and reproduction, by Rinia *et al.* (2001) for physics, by Morillo, Bordons and Gómez (2001) for chemistry and by Rinia *et al.* (2002) for all fields of science. Adams, Jackson and Marshall (2007) also use the proportion of cited references made to different disciplines, to which they add the number of distinct source categories cited as well as the Shannon Diversity Index.

On the other hand, Rinia, van Leeuwen and van Raan (2002) defined it as the % of papers from a group of researchers that is published outside their "main" discipline. For example, the degree of interdisciplinarity of physicists is the % of their papers published in journals outside the discipline of physics. Finally, Levitt and Thelwall (2008), in a study of the scientific impact of interdisciplinary research, defined interdisciplinarity articles as articles published in journals to which more than one discipline has been attributed (either by Thomson Reuters' Web of Science or by Elsevier's Scopus). This method was also used by Morillo, Bordons and Gómez (2003). Though this operationalization of interdisciplinarity (or of multidisciplinary) is simple to understand, the fact that a journal is attributed to more than one discipline does not imply that papers published in this journal are actually "interdisciplinary" papers. Such journal could be publishing disciplinary papers from different disciplines, without necessary having a dialog between the disciplines covered. This is the case, for example, of multidisciplinary journals like *Nature* and *Science*.

Our analysis uses all papers published in the year 2000 in journals covered by the Thomson Scientific's Web of Science, which includes the Science Citation Index Expanded (SCIE), the Social Sciences Citation Index (SSCI), and the Arts and Humanities Citation Index (AHCI). For each document indexed in Thomson's databases (source items), a list of references is included. This allows us, following Porter and Chubin (1985), to use the relationship between the disciplines of the cited and citing documents to measure the degree of interdisciplinarity of papers. The year 2000 was chosen because it allowed enough time for the items to be cited and thus permit the calculation of their impact. The disciplinary classification of journals used in this paper is that of the U.S. National Science Foundation NSF)[1]. This classification categorizes each journal into a single discipline and specialty. Since the NSF classification excludes arts and humanities, we categorized journals of the AHCI as belonging to either arts or humanities. Our classification includes 143 specialties which can be regrouped into 14 disciplines. Given the limits of bibliometric data for the measurement of the social sciences and, more importantly, arts and humanities (Larivière, Archambault, Gingras and Vignola-Gagné, 2006; Archambault, Vignola-Gagné, Côté, Larivière and Gingras, 2006), the trends observed for these domains must be interpreted with caution. We nonetheless included these two fields in order to measure the full spectrum of interdisciplinarity.

Rinia's (2007) Thesis presents two levels of interdisciplinarity: "big" and "small" interdisciplinarity. Big interdisciplinarity refers to interdisciplinarity occurring between different disciplines (e.g. chemistry and physics), while small interdisciplinarity refers to interdisciplinarity between different specialties (e.g. organic chemistry and applied chemistry). In this paper we limit the analysis to "interdisciplinarity", defined as relations between different disciplines and leave out "interspecialty" defined as links between different scientific specialties irrespective of their discipline. Thus, the degree or level of interdisciplinarity of a paper is defined as its percentage of references made to papers assigned to a discipline different from that of the citing paper. This percentage of course varies from 0% to 100%. For example, an article published in a chemistry journal that includes 12 references to papers published in chemistry journals and 8 references to journals in other disciplines (physics, clinical medicine, etc.)—for a total of 20 citations—obtains an interdisciplinarity index of 40 % (8/20).

A limitation of this method is that only references made to other source items in the database can be assigned to a given discipline. Globally this represents about 65% of all cited references, all disciplines combined. This percentage varies between disciplines and represents 79% of references in medical fields, 61% in the natural sciences, 37% in the social sciences and only 5% in arts and humanities. This large variation between disciplines is a reflection of the proportion of their references made to journal articles (Larivière, Archambault, Gingras and Vignola-Gagné, 2006). By construction, only papers with at least one reference made to WoS-indexed material are included in the study (N=750,743).

Scientific impact measures presented here are similar to those developed by Schubert and Braun (1986) and by Moed, De Bruin and van Leeuwen (1995). Hence, in order to take into account the fact that publication and citation practices vary according to disciplines, all impact measures are normalized by the world average of each specialty. Three measures of scientific impact are compiled: 1) average of relative citations (ARC) received by papers, 2) average of relative impact factor (ARIF) of journals in which the papers are published and 3) percentage of papers published in the top 5% most cited papers. In order to take into account the different aging patterns of papers and journals in the social sciences and in the humanities (Larivière,

---
[1] More details on the classification can be found at: http://www.nsf.gov/statistics/seind06/c5/c5s3.htm#sb1

Archambault and Gingras, 2008) the ARC of papers is calculated using a 5-year citation window following publication year and exclude first author self-citations. In the calculation of the impact factors, the asymmetry between the numerator and the denominator has been corrected. ARIF and ARC measures above (or below) one mean that they are above (or below) the world average in their respective discipline.

**Results**

Figure 1 presents, for three broad disciplinary categories, the distribution of papers by (rounded[2]) percentage of interdisciplinary references. One can readily see that a significant share of the papers—one third in both social sciences and natural sciences and medicine and two-thirds in arts and humanities—are essentially disciplinary (<5% of references made to other disciplines). The very low level of interdisciplinarity of the arts and humanities is consistent with observations of Morillo, Bordons and Gómez (2003). In social sciences and in arts and humanities, we see that both extremes have a high number of papers while the intermediate levels between 5% and 95% are about equally distributed. Among the fields included in social sciences, only health-related papers follow a different trend: the number of papers increases linearly with the percentage of interdisciplinarity. In natural sciences and medicine, however, we see a continuous decrease in the proportion of papers with the rise of the level of interdisciplinarity, followed by a slight increase for the highest level of interdisciplinarity (>95% of references outside the paper's discipline). Globally, the majority of papers have a low level of interdisciplinarity and a minority of papers has a high score on this interdisciplinarity index.

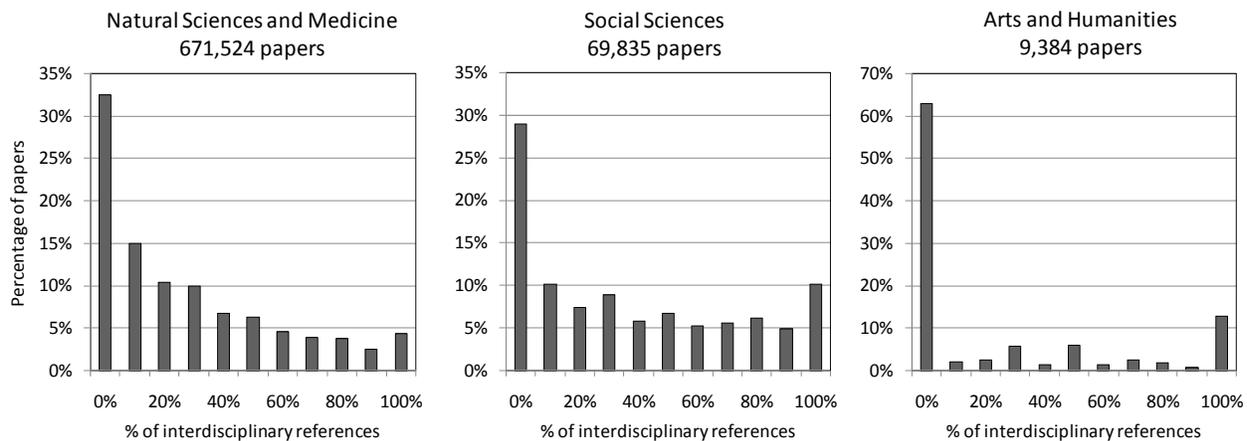

Figure 1. Distribution of papers by percentage of references made to journals of other disciplines, by broad discipline, 2000

Figure 2 and 3 presents, for social sciences and humanities and natural sciences and medicine, the relationship between the interdisciplinarity of references and the scientific impact of papers—average of relative citations of papers (ARC), average of relative impact factor (ARIF) and percentage of papers in the top 5% papers. Most histograms show that the highest levels of disciplinarity and interdisciplinarity have significantly lower impact scores than those in the middle range. This characteristic is observed in all disciplines: purely disciplinary (<5%) and purely interdisciplinary papers (>95%) obtain, on average, lower citation rates, are published in lower impact factor journals and are less likely to be amongst the 5% most cited papers. This suggests that papers that are either too disciplinary or too interdisciplinary are perhaps too

---
[2] Percentages between 0 and 4.99 are compiled as 0, those between 5 and 14.99 as 10, etc

mainstream or too much dispersed and, hence, do not attract as much attention as papers with more balanced mix of cited papers from different disciplines.

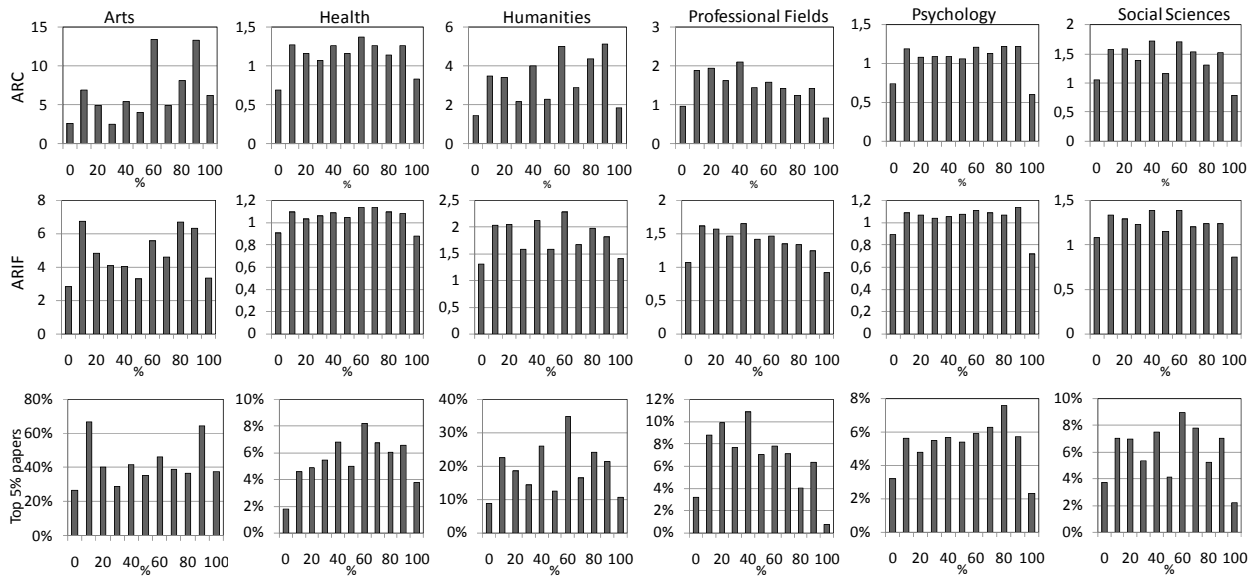

Figure 2. Distribution of the average of relative citations of papers (ARC), average of relative impact factor (ARIF) and percentage of papers in the top 5% papers, by percentage of references made to journals of other disciplines, for disciplines of social sciences and humanities, 2000

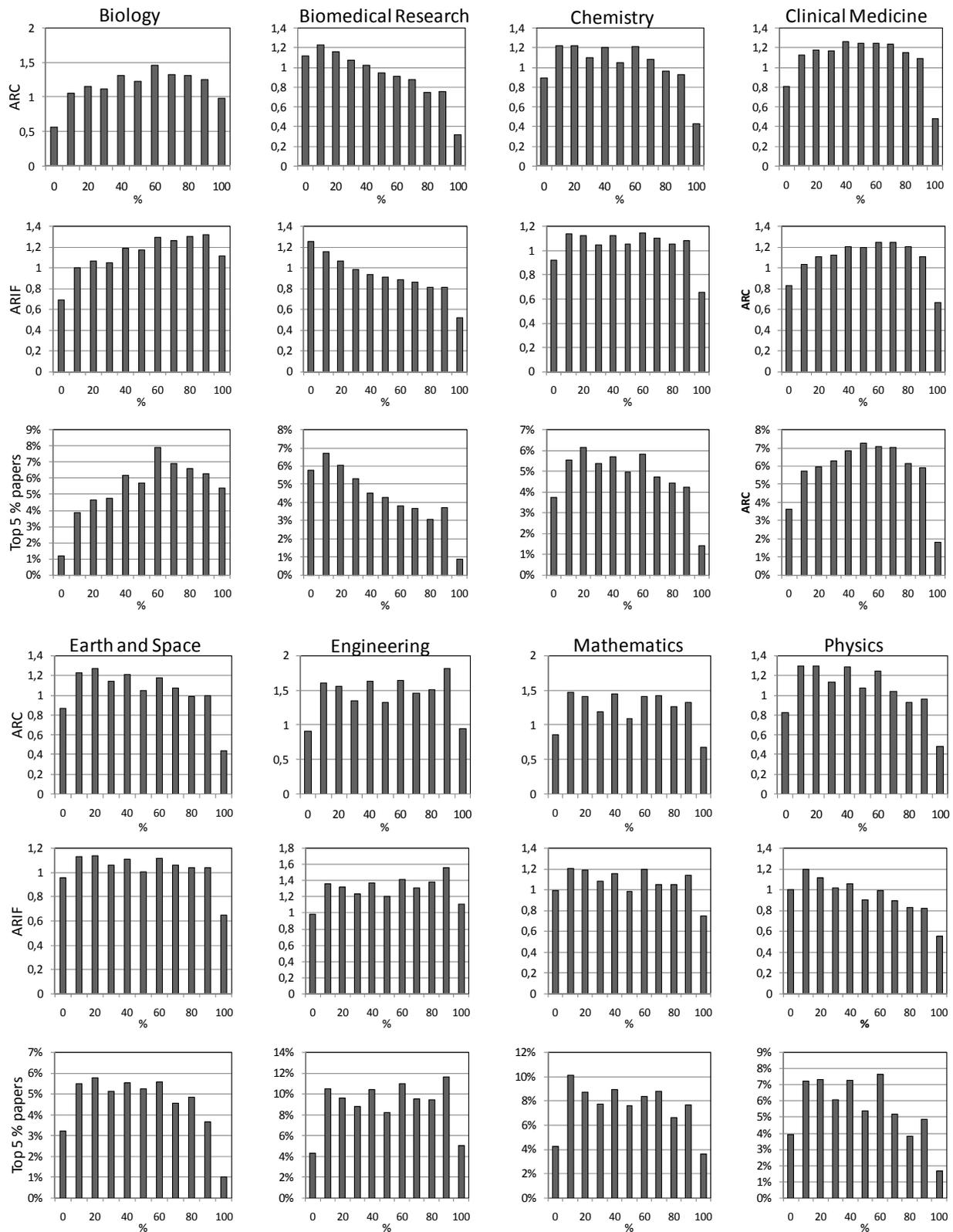

Figure 3. Distribution of the average of relative citations of papers (ARC), average of relative impact factor

(ARIF) and percentage of papers in the top 5% papers, by percentage of references made to journals of other disciplines, for disciplines of natural sciences and engineering, 2000

Apart from this common characteristic of the histograms, two different patterns emerge from the data; each of them corresponding to a different group of disciplines. The first pattern, exemplified by the disciplines of biomedical research, earth and space science, physics and professional fields is that scientific impact is negatively correlated with higher interdisciplinarity. In other words, in these disciplines, papers having more references to articles published in journals belonging to other disciplines obtain, on average, lower impact scores. Among these disciplines, biomedical research is the discipline in which this relation is the most obvious, with $r^2$ values of 0.82 for ARC, 0.88 for ARIF and 0.89 for the top 5% papers. The second pattern is that higher interdisciplinarity is linked with an increase of the scientific impact of papers—until it reaches a plateau at about 60% of references made to journals of other disciplines. This is the case of biology, clinical medicine, health (social sciences), humanities and psychology. Finally, for the disciplines of arts, chemistry, engineering, mathematics and social sciences, only extremes values of interdisciplinarity (<5% and >95% of outside discipline references) are lower, but no distinct pattern can be seen in the middle-range values.

**Discussion and Conclusion**
Though it is often implicitly suggested that being more interdisciplinary is inherently a good thing for research, such a conclusion is rarely based on solid empirical data and constitute more a wish than a tested fact. Instead of taking the level of interdisciplinarity as a definition of "good research", we have measured the relation between the level of interdisciplinarity of individual papers—defined through the disciplinary structure of their references—and their scientific impact, using three indicators (field-normalized citation rates, impact factor, and percentage of top 5% most cited papers).

Our results show that, although there is no clear correlation for all disciplines combined between the degree of interdisciplinarity of papers and their citation scores, there are nonetheless some disciplines in which higher levels of interdisciplinarity are linked with higher citation rates, and other where high levels of interdisciplinarity correlate with a lower citation rate. One characteristic is, however, observed in all disciplines: the highest levels of disciplinarity as well as the highest levels of interdisciplinarity (respectively <5% and >95% of references to other disciplines) have lower scientific impact than the papers whose levels of interdisciplinarity is in between those extremes. This suggests that there might be an optimum of interdisciplinarity beyond which the research is too dispersed to find its niche and under which it is too mainstream to have high scientific impact. In biomedical research and, to a lesser extent, in physics, earth and space sciences and professional fields, a higher degree of interdisciplinary is correlated with lower citation rates. On the other hand, in biology, clinical medicine, humanities, psychology and health (social sciences), a moderate interdisciplinarity is associated with higher citation rates.

The difference between disciplines in the relation linking interdisciplinarity and scientific impact could be related to the characteristics of the disciplines cited. For instance, these differences could be explained by the fact that some disciplines (like biomedical research and clinical medicine) are more citation intensive than mathematics or engineering (Wallace, Larivière and Gingras, 2009). Hence, papers having more interdisciplinary linkages with those fields might be more cited by these disciplines. We tested this hypothesis using all physics papers published in 2000 (Table 1) and found that, indeed, physics papers citing more than 50% of biomedical research and clinical medicine papers had statistically significant higher ARC values than physics papers citing less than 50% of papers from such disciplines. The opposite phenomenon is observed for all other disciplines—except, of course, physics itself—but is more striking for non citation intensive disciplines like mathematics and engineering.

Table 1. Average of relative citations of papers (ARC) of physics papers, by percentage of references made to papers from other disciplines and mean number of citations received by all papers in each discipline, 2000

| Discipline | % of references made to each discipline | | Mean number of citations per paper |
| --- | --- | --- | --- |
| | Below 50% | 50% and above | |
| Biomedical Research | 1.03 | 1.26 | 28.57 |
| Clinical Medicine | 1.03 | 1.14 | 17.01 |
| Earth and Space Sciences | 1.03 | 0.90 | 13.57 |
| Chemistry | 1.04 | 0.84 | 12.84 |
| Biology | 1.03 | 0.69 | 10.52 |
| Physics | 0.9 | 1.05 | 10.15 |
| Engineering | 1.05 | 0.68 | 5.68 |
| Mathematics | 1.03 | 0.54 | 4.05 |

In order to provide more insight into this relationship, we also measured the correlation, for each physics paper, between the percentage of its references made to a discipline and the percentage of citations received from this discipline. More precisely, is a physics paper for which 35% of the references are made to articles of biomedical research receiving a similar percentage of its citations from this discipline? Unsurprisingly, physics papers having a higher share of their references made to biomedical research articles were more likely to receive citations from biomedical research articles. This correlation was strong for all disciplines of the natural sciences and medicine, with Pearson's $r$ between 0.50 and 0.63. This clearly shows that the relationship between interdisciplinarity and scientific impact is highly determined by the citation characteristics of the disciplines involved, as papers citing citation intensive disciplines are more likely to be cited by those disciplines and, hence, obtain higher citation rates than papers citing non citation intensive disciplines. This fact has important consequences on the interpretation given to the relation between citations and interdisciplinarity as it shows that higher citation is linked to the citation intensive disciplines and not necessarily to the intrinsic quality of the paper itself. Thus, a strictly rigorous measure of the link between citation and quality of papers should compare papers with the same composition of interdisciplinary references.

**Acknowledgments**
The authors wish to Jean-Pierre Robitaille, Ismael Rafols, as well as the two anonymous referees for their useful comments and suggestions.